\documentclass[aps,twocolumn,amssymb,amsfonts,amsmath,showpacs,final,a4paper,superscriptaddress]{revtex4-1}

\usepackage{float}
\usepackage[dvips,final]{graphicx}
\usepackage{color}
\usepackage{amsmath}

\begin{document}

\title{Local and non-local electron dynamics of Au/Fe/MgO(001) heterostructures analyzed by time-resolved two-photon photoemission spectroscopy}

\author{Y. Beyazit}
\affiliation{Faculty of Physics and Center for Nanointegration (CENIDE), University of Duisburg-Essen, Lotharstr.~1, 47057 Duisburg, Germany}

\author{J. Beckord}
\altaffiliation[]{Current address: University of Zurich, 8057 Zurich, Switzerland}
\affiliation{Faculty of Physics and Center for Nanointegration (CENIDE), University of Duisburg-Essen, Lotharstr.~1, 47057 Duisburg, Germany}

\author{P. Zhou}
\affiliation{Faculty of Physics and Center for Nanointegration (CENIDE), University of Duisburg-Essen, Lotharstr.~1, 47057 Duisburg, Germany}

\author{J. P. Meyburg}
\affiliation{Faculty of Chemistry, University of Duisburg-Essen, Universit\"{a}tsstr.~5, 45141 Essen, Germany}

\author{F. K\"{u}hne}
\affiliation{Faculty of Chemistry, University of Duisburg-Essen, Universit\"{a}tsstr.~5, 45141 Essen, Germany}

\author{D. Diesing}
\affiliation{Faculty of Chemistry, University of Duisburg-Essen, Universit\"{a}tsstr.~5, 45141 Essen, Germany}

\author{M. Ligges}
\altaffiliation{Current address: Fraunhofer IMS, 47057 Duisburg, Germany}
\affiliation{Faculty of Physics and Center for Nanointegration (CENIDE), University of Duisburg-Essen, Lotharstr.~1, 47057 Duisburg, Germany}

\author{U. Bovensiepen}
\email[]{uwe.bovensiepen@uni-due.de}
\affiliation{Faculty of Physics and Center for Nanointegration (CENIDE), University of Duisburg-Essen, Lotharstr.~1, 47057 Duisburg, Germany}

\date{\today}

\begin{abstract}

Employing femtosecond laser pulses in front and back side pumping of Au/Fe/MgO(001) combined with detection in two-photon photoelectron emission spectroscopy we analyze local relaxation dynamics of excited electrons in buried Fe, injection into Au across the Fe-Au interface, and electron transport across the Au layer at 0.6 to 2.0~eV above the Fermi energy. By analysis as a function of Au film thickness we obtain the electron lifetimes of bulk Au and Fe and distinguish the relaxation in the heterostructure's constituents. We also show that the excited electrons propagate through Au in a superdiffusive regime and conclude further that electron injection across the epitaxial interface proceeds ballistically by electron wavepacket propagation.

\end{abstract}

\maketitle

Excited charge carriers relax in metals and semiconductors on femto- to picosecond timescales due to the large  phase space for electron-electron (e-e) and electron-phonon scattering \cite{Shah99, Bauer15}. Microscopic insight into these processes was developed by combined efforts of static spectroscopy, spectroscopy in the time domain, and \emph{ab initio} theory \cite{ECHENIQUE2004}. Early optical experiments used back side pump / front side probe schemes and analyzed the propagation dynamics through the bulk of thin films \cite{Brorson1987}. Time- and angle-resolved two-photon photoelectron spectroscopy exploited the sensitivity to electron energy and momentum and was key to develop a comprehensive understanding of the microscopic nature of the engaged elementary processes which hot electrons experience \cite{harris_local_1998, weinelt2004, Guedde2007, Bovensiepen12, Cui2014}. In heterostrutures such an analysis is challenging but highly desired given the widespread application of these material systems. The surface sensitivity of photoelectron spectroscopy is a severe limitation for heterostructures and buried media which can be overcome by using hard X-ray photons in photoemission \cite{Woicik16,oloff14}. Also photoelectrons with low kinetic energy are reported to probe the bulk \cite{kiss_05} or buried interface electronic structure \cite{rohleder_05,rettig12} in selected cases.

Hot electrons are characterized by their energy above the Fermi energy $E-E_{\mathrm{F}}\gg k_{\mathrm{B}}T$, $T$ is the equilibrium temperature, and their momentum $\mathbf{k}$. For a component $k_\perp$ directed from the surface into bulk, transport effects occur. So far, local dynamics at the surface and non-local contributions due to, e.g., transport were distinguished indirectly by analyzing relaxation at surfaces \cite{aeschlimann_APA00,liso_APA04,liso_APA04b,GAUYACQ2007,kirchmann_NatPhys10}. Particular systems allowed a microscopic description of electron propagation through a molecular layer \cite{staehler_08} and resonant tunneling across a dielectric film \cite{rohleder_05}.

Electronic transport properties are essential in condensed matter. Besides transport in Bloch bands at $E_{\mathrm{F}}$ problems like incoherent hopping in molecular wires \cite{kocherzhenko_09} and two dimensional materials \cite{Ngankeu17,lee_19}, superdiffusive spin currents \cite{Battiato2010}, and attosecond phenomena at surfaces \cite{Tao16} are important. The relevance of spin-dependent charge carrier transport in femtosecond magnetization dynamics has spurred the use of back side pumping in optical pump-probe experiments \cite{Melnikov2011, bergeard_16, Razdolski2017a}, which provide energy and momentum integrated information. Also detection in microscopes provides insight into carrier propagation \cite{Sung2020} and plasmon dynamics \cite{klick2019}. Back side pump / front side probe photoelectron spectroscopy might have considerably impact, since it promises energy and momentum dependent information \cite{Sung2020}.

In this letter we report such a back side pump / front side probe photoemission experiment for the model system Au/Fe/MgO(001). While for thin films the electronic relaxation agrees for front and back side pumping, we identify electron transport for thicker Au films upon back side pumping. We are aware that the dynamics has a spin-dependent contribution \cite{Melnikov2011}. The experiment performed here is spin-integrating and we focus on the charge dynamics. By analyzing the relaxation time dependence on the Au film thickness $d_{\mathrm{Au}}$ we distinguish the electron dynamics in the Au and Fe constituents.

\begin{figure}
    \centering
        \includegraphics[width=0.85\columnwidth]{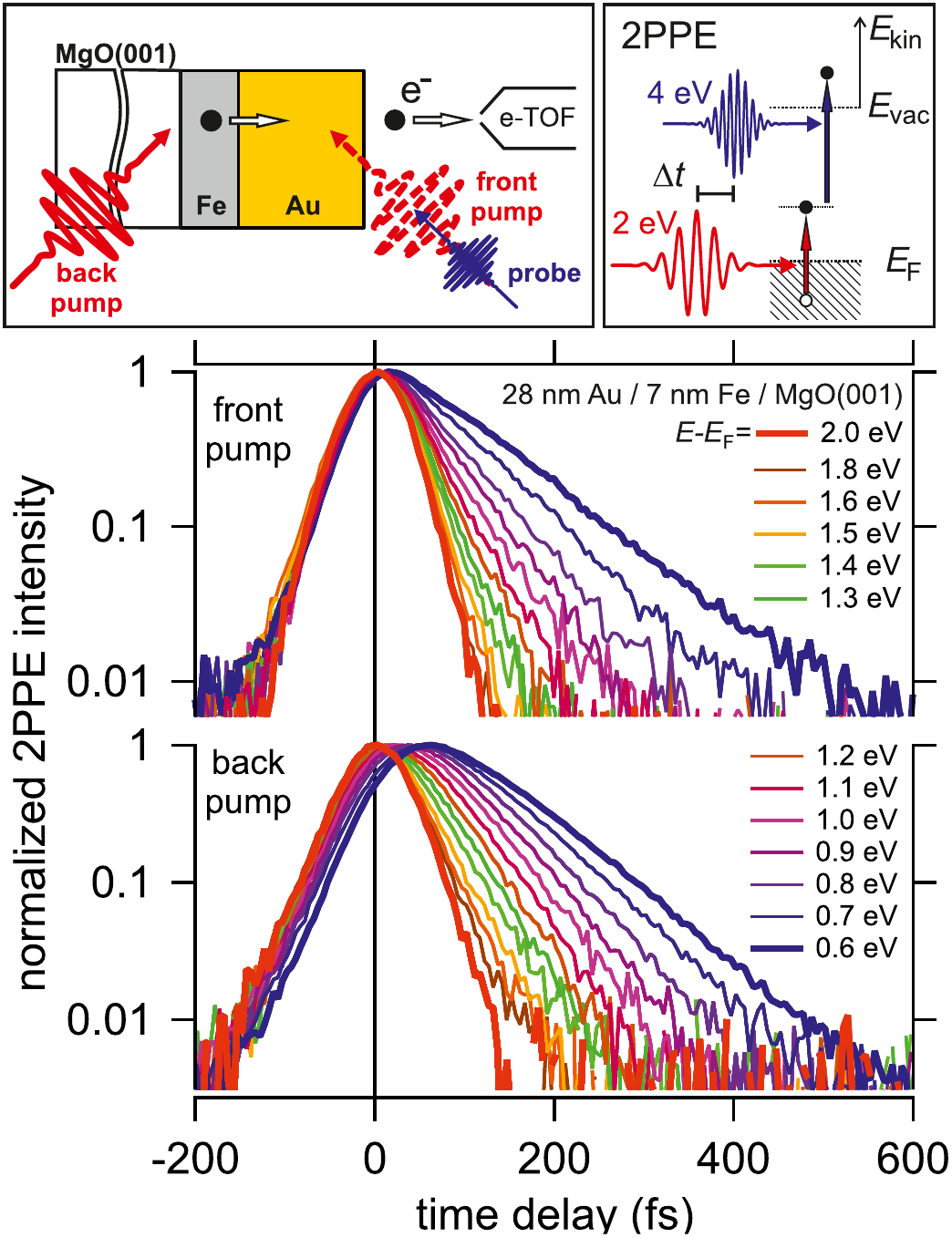}
\caption{Top: Schematic experimental configuration for front / back side pumping depicting the photoelectron analysis and the two-color 2PPE process for absorption of one 2~eV and 4~eV photon each. Bottom: Time-dependent 2PPE intensity on a logarithmic scale for front (top) and back side pumping (bottom) at indicated energies above the Fermi level.}
    \label{fig:1}
\end{figure}

Fig.~\ref{fig:1}, top, depicts the experimental configuration. Femtosecond laser pulses are generated by a commercial regenerative Ti:sapphire amplifier (Coherent RegA 9040) combined with a non-collinear optical parametric amplifier (NOPA, Clark-MXR) operating at 250~kHz repetition rate. We use pairs of 2 and 4~eV pulses each of 50~fs pulse duration as pump and probe pulses, respectively. Pump pulses are sent to the Au/Fe/MgO(001) sample kept at room temperature with $\pm45^{\circ}$ angle of incidence, see Fig.~\ref{fig:1}, and reach at first the Au surface in case of front side pumping or are transmitted through MgO(001) and excite electrons of Fe in back side pumping. The incident pump fluence was 50~$\mu$J/cm$^2$. Probe pulses are in both configurations sent to the Au surface close to $45^{\circ}$ and probe electrons in intermediate, excited states (i) at the surface in case of front side pumping or (ii) propagating through Au for back side pumping. The time-delayed probe pulse photoemitts electrons from the excited state by absorption of one 4~eV photon, see Fig.~\ref{fig:1}, top right, and generates the two-color two-photon photoemission (2PPE) signal discussed here. Photoelectrons are analyzed by a time-of-flight (e-TOF) analyzer \cite{kirchmann_APA08} and collected within $\pm11^{\circ}$ off the surface normal, see Fig.~\ref{fig:1}. For a discussion of one- and two-color 2PPE, see supplement \cite{suppl}. The Au/Fe/MgO(001) system was chosen due to its epitaxial structure \cite{rickart2001} and inert surface. Samples are grown by molecular beam epitaxy, stored under Ar atmosphere, transferred in ambient conditions to the photoemission chamber, and cleaned by heating to $80^{\circ}$C. Crystalline order and layer thickness were analyzed on a twin sample by CROSS-TEM, magnetometry, profilometry, and AFM, see \cite{suppl,Razdolski2017a,Alekhin2017}.

Fig.~\ref{fig:1} shows the time-dependent 2PPE intensity for back and front side pumping at selected energies $E-E_{\mathrm{F}}$ for 28~nm~Au/7~nm~Fe/MgO(001). Time zero is defined by the fastest signal given by the intensity maximum of electrons at the maximum kinetic energy, see $E-E_{\mathrm{F}}=2.0$~eV in Fig.~\ref{fig:1} \cite{ball}. The upper graph depicts front side pump 2PPE data. The lower panel shows results for back side pumping which exhibit a shift in time delay of the intensity built-up and maximum increasing with decreasing $E-E_{\mathrm{F}}$. This effect is assigned to delayed arrival of excited electrons at the Au-vacuum interface. Both data sets exhibit slower intensity relaxation for lower $E-E_{\mathrm{F}}$ due to the respective increase in hot electron lifetime \cite{Bauer15}.

\begin{figure}
    \centering
        \includegraphics[width=0.9\columnwidth]{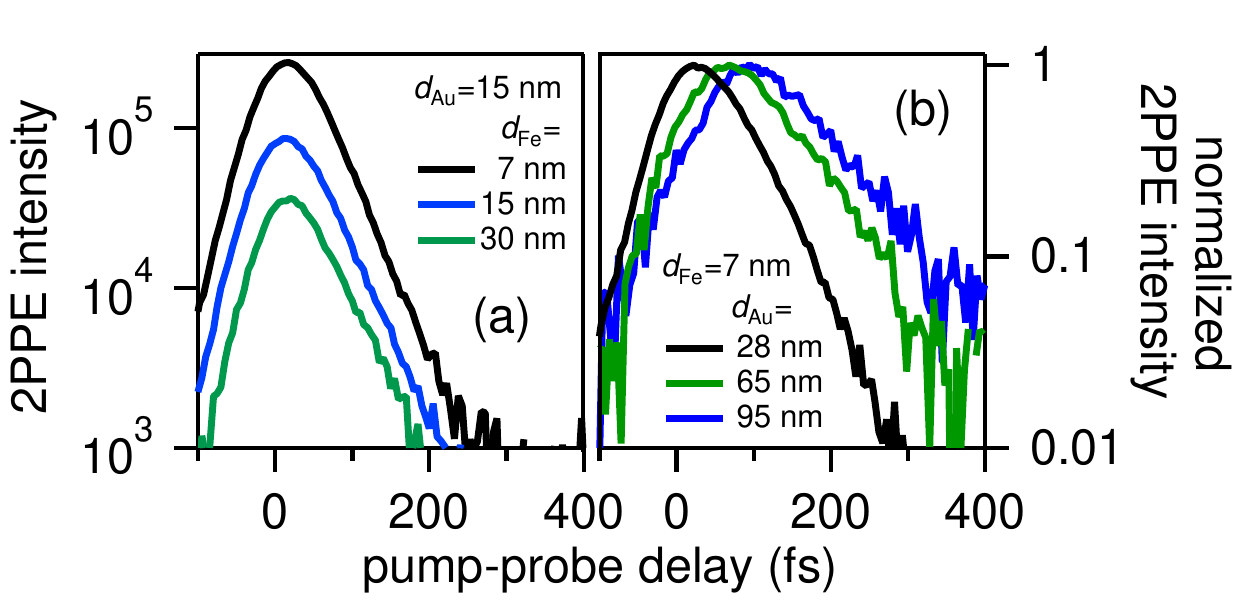}
\caption{Time-dependent 2PPE intensity upon back side pumping at $E-E_{\mathrm{F}}=$1.0~eV shown in (a) for different $d_{\mathrm{Fe}}$ and constant $d_{\mathrm{Au}}$ and vice versa in (b). In the latter case the 2PPE intensity is normalized to the intensity maximum.}
    \label{fig:2}
\end{figure}

Fig.~\ref{fig:2} compares back side pump 2PPE for different Fe thickness $d_{\mathrm{Fe}}$ (a) and for different $d_{\mathrm{Au}}$ (b). While in case of larger $d_{\mathrm{Au}}$ transport effects are identified through a time shift in arrival at the Au surface, increasing of $d_{\mathrm{Fe}}$ results essentially in a loss of intensity. Note that such loss of intensity is also observed with increasing $d_{\mathrm{Au}}$, see supplementary material \cite{suppl}, because only electrons which reach the Au-vacuum interface are detected. These observations support the following concept. The Fe layer acts as the optically excited electron emitter and the Au layer serves as the acceptor hosting electron propagation as depicted by the scheme in Fig.~\ref{fig:1}.

\begin{figure}
    \centering
        \includegraphics[width=0.85\columnwidth]{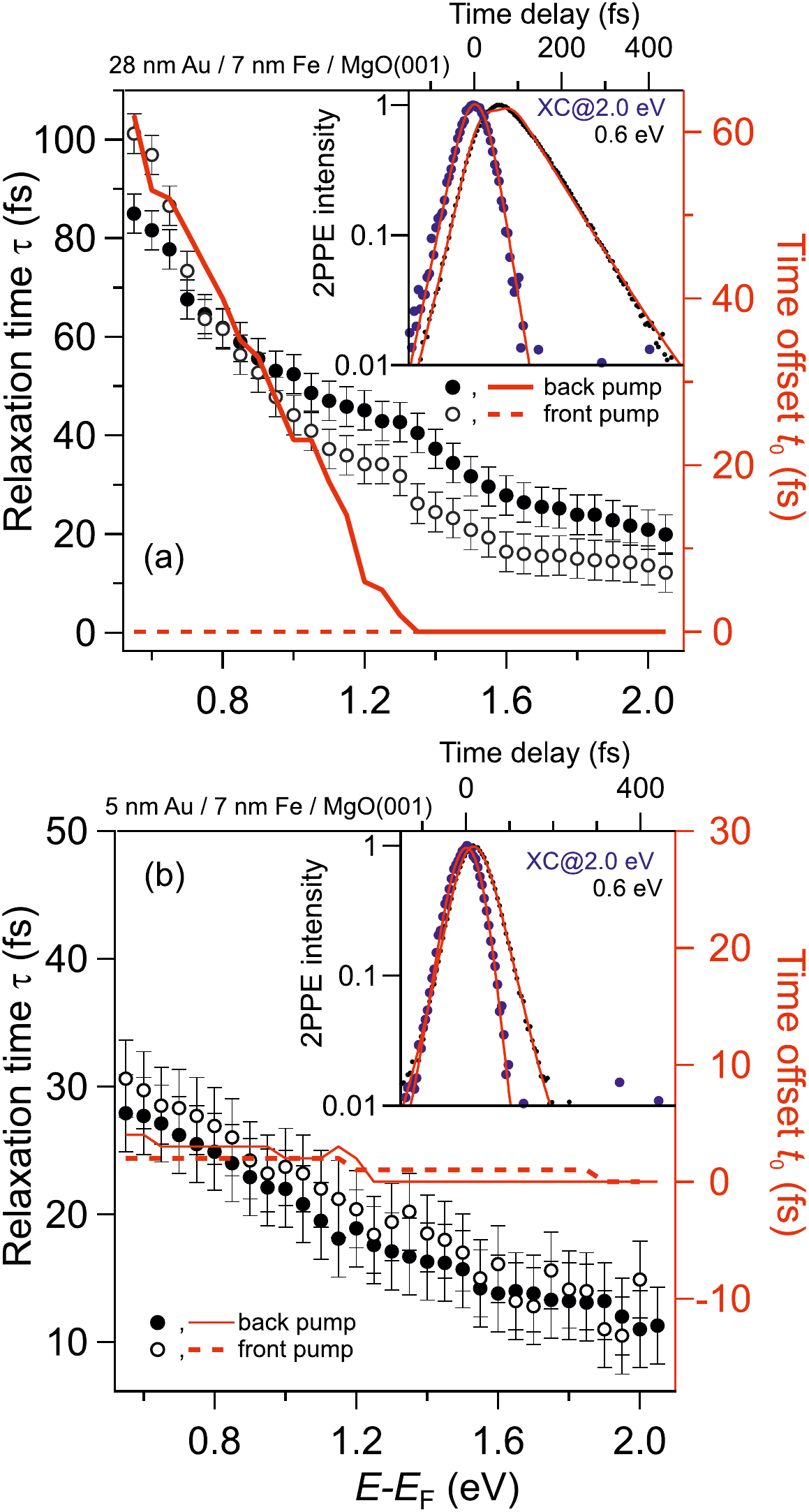}
\caption{Left axis: relaxation times $\tau$ of hot electrons at energies $E-E_{\mathrm{F}}$ for 28~nm (a) and 5~nm (b) thick Au films on 7~nm Fe on a MgO(001) substrate. Right axis: Time offset $t_0$, see text. Both quantities are determined by fitting and are given for front and back side pumping as indicated. Error bars for $t_0$ are $\pm$6~fs at 0.7~eV and decrease to $\pm$3~fs at 2.0~eV.}
    \label{fig:3}
\end{figure}

Time-dependent 2PPE intensities are fitted by a single exponential decay $\propto \exp{\frac{t-t_0}{\tau}}$ convolved with the cross correlation (XC) of the laser pulses as determined by 2PPE at maximum kinetic energy. Examples of such fits are plotted as insets in Fig.~\ref{fig:3}(a,b). This fitting determines energy dependent, inelastic relaxation times $\tau(E)$ and time offsets $t_0(E)$, at which the relaxation starts. Fig.~\ref{fig:3} shows $\tau(E)$ and $t_0(E)$ obtained for $d_{\mathrm{Au}}$=28~nm (a) and 5~nm (b) at $d_{\mathrm{Fe}}$=7~nm. We find a decrease in $\tau$ with increasing energy and -- if compared at identical energies -- $\tau$ is larger for the thicker than for the thinner Au layer. Front and back side pumping lead to small differences in $\tau(E)$ near 1.2~eV for $d_{\mathrm{Au}}$=28~nm. Such differences were not obtained for $d_{\mathrm{Au}}$=5~nm, neither in $\tau$ nor in $t_0$. For sufficiently thin films transport effects become negligible \cite{liso_APA04b} and $d_{\mathrm{Au}}$=5~nm provides a reasonable reference value in this regard. For $d_{\mathrm{Au}}$=28~nm we identify for back side pumping variations in $t_0(E)$ up to 60~fs while $t_0(E)$ does not vary for front side pumping. The observed $t_0>0$ for back side pumping is a non-local effect and quantifies the time required for transport of electrons excited in Fe, injected into Au across the Fe-Au interface, and propagate through Au towards the Au-vacuum interface, where they are probed in 2PPE. Following bulk optical constants \cite{weaver81} 95\% of the absorbed pump pulse intensity excites the 7~nm Fe layer and the 2PPE signal detected for 28~nm Au / 7~nm Fe / MgO(001) is dominated by electrons propagating through Au. This assignment is supported by the increase in $t_0$ and $\tau$ for the larger $d_{\mathrm{Au}}$ compared to the thinner one, see Fig.~\ref{fig:3}. Since relaxation times of hot electrons in metals at few eV energy above $E_{\mathrm{F}}$ are determined by inelastic e-e scattering \cite{Bauer15}, the similar trend of increasing $t_0$ and $\tau$ with decreasing energy indicates that $t_0$ is determined by inelastic e-e scattering as well. On this basis the electron transport through Au is concluded to proceed in a superdiffusive regime, which occurs before hot electrons have thermalized by subsequent e-e scattering events \cite{Battiato2012}. As discussed in the supplement \cite{suppl}, reaching the limit of diffusion would require many scattering processes, which we exclude due to the observation $t_0(E_0)<\tau(E_0)$, which implies individual events. Ballistic propagation, on the other hand, would occur for absent relaxation, which disagrees with the observed temporal broadening in time-dependent 2PPE intensities while the electrons propagate through Au, see Fig.~\ref{fig:2}(b). Given the weak variation of the electron group velocity with respect to the Fermi velocity in Au \cite{nenno_18} ballistic propagation is also incompatible with the increase in $t_0$ observed with decreasing energy. Scattering might increase the covered distance to the surface and a determination of the electron's propagation velocity $v=d_{\mathrm{Au}}/t_0$, which results for $d_{\mathrm{Au}}$=28~nm at $E-E_{\mathrm{F}}$=1.0~eV in $v$=1.3~nm/fs -- a value close to $v_{\mathrm{F}}$ in Au -- has to be treated with care. We note that we cannot exclude ballistic propagation of electrons $E-E_{\mathrm{F}}\geq$1.3~eV where we find $t_0=0$~fs, which is set by the time zero determination.

\begin{figure}
    \centering
        \includegraphics[width=\columnwidth]{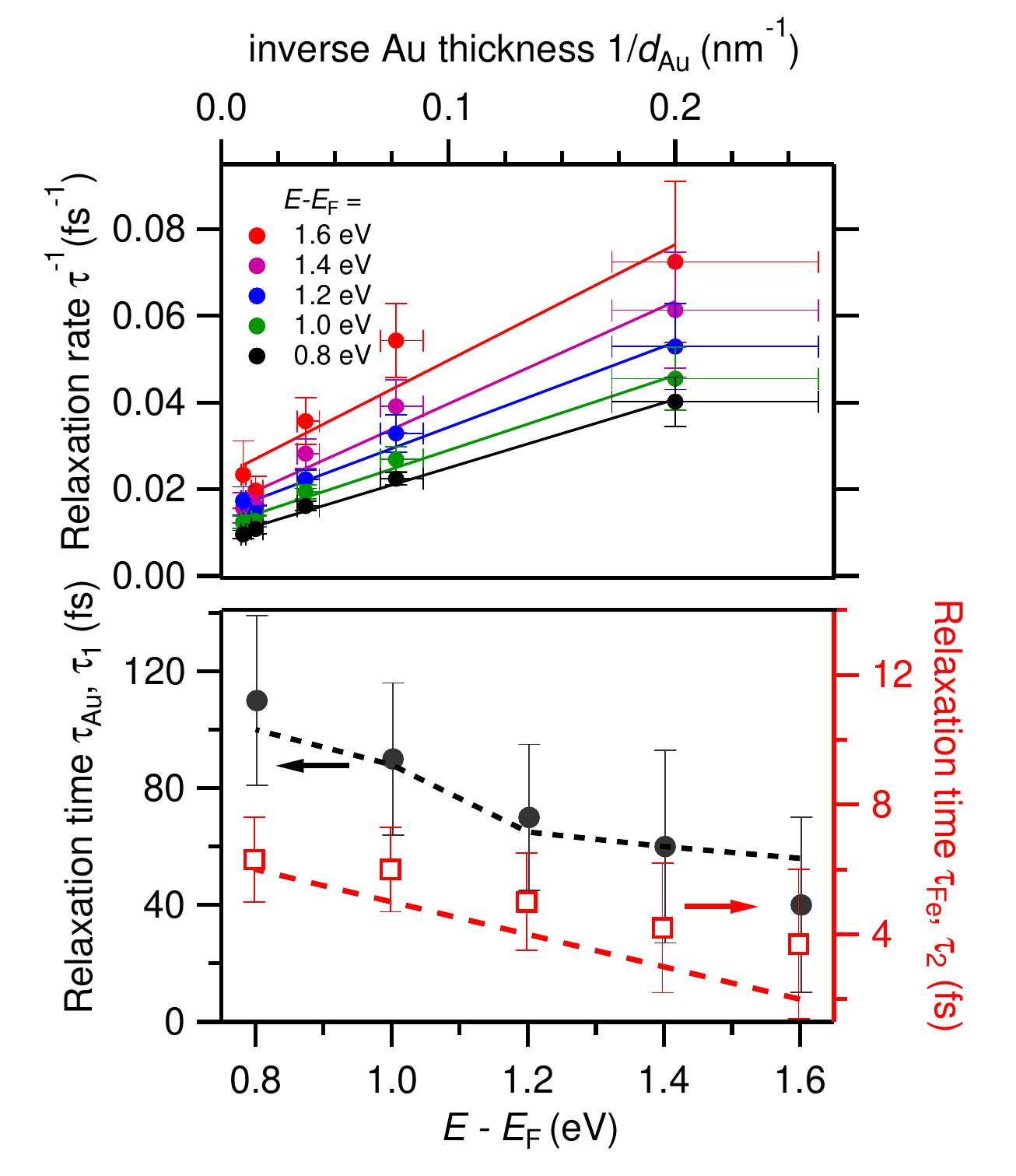}
\caption{Top: Relaxation rates $\tau^{-1}$ as a function of $d_{\mathrm{Au}}^{-1}$ for different electron energies above $E_{\mathrm{F}}$. Lines represent linear fits. Slopes and intercepts determine the two relaxation rates discussed in the text. Bottom panel: The determined relaxation times $\tau_1$ ($\bullet$) and $\tau_2$ ($\square$) as a function of $E-E_{\mathrm{F}}$ in comparison with literature data for hot electron lifetimes $\tau_{\mathrm{Au}}$ and $\tau_{\mathrm{Fe}}$ from Ref.~\cite{Bauer15} shown by dashed lines.}
    \label{fig:4}
\end{figure}

We investigate $d_{\mathrm{Au}}=5-95$~nm and identify a thickness dependent $\tau=\tau(d_{\mathrm{Au}})$, see Figs.~\ref{fig:2}, \ref{fig:3}, which is for thinner films smaller than in bulk Au \cite{Bauer15}. Fig.~\ref{fig:4}, top, shows $\tau(d_{\mathrm{Au}})^{-1}$ for different energy. To understand this thickness dependence we consider a continuum approach to scattering in the heterostructure, which assumes that the individual thicknesses $d_{\mathrm{Au}}$, $d_{\mathrm{Fe}}$, and the extension of the interface $d_{\mathrm{Au-Fe}}$ are comparable with the respective scattering lengths $\lambda_i \approx \tau_i \cdot v_{\mathrm{F},i}$, which are $\approx 50$~nm in Au and $\approx 2$~nm in Fe \cite{Zhukov2006,Razdolski2017a}. The integral scattering probability of electrons propagating in the interface normal direction $z$ increases linearly with $d_{\mathrm{Au}}$, $d_{\mathrm{Fe}}$, and $d_{\mathrm{Au-Fe}}$:

\begin{equation}
\int \limits_{0}^{d_{\mathrm{Fe}}+d_{\mathrm{Au-Fe}}+d_{\mathrm{Au}}}\frac{dz}{\tau(z)} =\frac{d_{\mathrm{Fe}}}{\tau_{\mathrm{Fe}}}+\frac{d_{\mathrm{Au-Fe}}}{\tau_{\mathrm{Au-Fe}}}+\frac{d_{\mathrm{Au}}}{\tau_{\mathrm{Au}}}.
\label{eq:1}
\end{equation}

In our 2PPE back side pump -- front side probe experiment the variation of $d_{\mathrm{Au}}$ allows separation of two independent processes, see supplement \cite{suppl}, described by

\begin{equation}
\frac{1}{\tau(d_{\mathrm{Au}})}=\frac{1}{\tau_{\mathrm{1}}}+\frac{1}{\tau_{\mathrm{2}}}=A+\frac{B}{d_{\mathrm{Au}}}.
\label{eq:2}
\end{equation}

Fig.~\ref{fig:4}, top, depicts fits following Eq.~\ref{eq:2} with $A$ and $B$ being the intercept with the ordinate and slope as a function of $1/d_{\mathrm{Au}}$, respectively. Note that both $A$ and $B/d$ have the dimension of a rate. Our analysis determines relaxation times $\tau_{1}$ and $\tau_{2}$ which are plotted in Fig.~\ref{fig:4}, bottom, in comparison with literature values for hot electron lifetimes in bulk Au and Fe $\tau_{\mathrm{Au}}, \tau_{\mathrm{Fe}}$, respectively, taken from Ref.~\cite{Bauer15}. Given the agreement of our data with these values, we conclude to have distinguished the electron dynamics in the two constituents and thereby demonstrate sensitivity to the buried Fe film.

The observation of $\tau{_{\mathrm{Au}}}$ is straightforward to understand. A hot electron injected into Au at the Fe-Au interface propagates through the Au film and reaches the surface where it is photoemitted. During the propagation it experiences inelastic e-e scattering with rates of bulk Au and transfers energy to a secondary electron. We detect this scattering by the time-dependent reduction of 2PPE intensity at the energy of interest at which the electrons were injected into Au. We did not take secondary electrons into account because we restricted the energy scale to rather high values $E-E_{\mathrm{F}}$, where primary electrons dominate \cite{liso_APA04b}. Secondary electrons start to contribute at half the primary energy \cite{Bauer15}, which is $E-E_{\mathrm{F}}<1$~eV for the highest energy electrons at 2~eV studied here. For sufficiently thick Au films, the second term in Eq.~\ref{eq:2} vanishes and scattering in Au dominates.

Understanding the determination of $\tau_{\mathrm{Fe}}$ in buried Fe requires consideration of all processes that may contribute to $B$ in Eq.~\ref{eq:2}. Following Eq.~\ref{eq:1} we take scattering in Fe and at the Au-Fe interface into account. Since we find within the experimental uncertainty $\tau_2=\tau_{\mathrm{Fe}}$ we conclude that the scattering at the interface does not contribute. In the investigated epitaxial heterostructure electron injection across the interface can be assumed to proceed by coherent propagation of a wavepaket in Bloch states which conserves energy and momentum across the interface \cite{GAUYACQ2007,Alekhin2017}. Therefore, the injection process across the Au-Fe interface is ballistic and violates our above assumption $d_{\mathrm{Au-Fe}}\approx\lambda_{\mathrm{Au-Fe}}$, which might be reason for not detecting it.

In conclusion, we demonstrated a time-domain analysis of electron dynamics in epitaxial Au/Fe/MgO(001) heterostructures with a total thickness of 12 - 102~nm. We distinguish the energy-dependent scattering rates in Fe and Au using optical pumping of Fe and detection at the Au surface by two-photon photoemission. We also identify the electron propagation to proceed in a superdiffusive regime. This separation of electron dynamics in the individual heterostructure constituents showcases the impact our approach might have on future work. A spectroscopy which accesses buried interfaces / media and provides energy-dependent information on electron dynamics is rarely available but may provide highly desired insights into heterostructures in general. We expect that this approach will bridge conventional transport measurements and time domain spectroscopy and will facilitate a more comprehensive understanding of electron dynamics in complex materials. We expect further that this approach to electron transport dynamics will be applied to semiconducting or insulating material systems due to its sensitivity to excited electronic states.

\begin{acknowledgments}
We acknowledge A. Eschenlohr for fruitful discussions and S. Salamon for experimental support. This work was funded by the Deutsche Forschungsgemeinschaft (DFG, German Research Foundation) Projektnummer  278162697 - SFB 1242.
\end{acknowledgments}

\providecommand{\noopsort}[1]{}\providecommand{\singleletter}[1]{#1}%

\end{document}